\documentclass[12pt]{article}
\usepackage{amsfonts}
\usepackage{amsmath}
\usepackage{graphicx}
\usepackage{amssymb}

\numberwithin{equation}{section}

\newtheorem{theorem}{Theorem}[section]

\newtheorem{definition}{Definition}

\def\bom{{\mbox{\boldmath$\omega$}}}
\def\obom{\overline\bom}
\def\0bom{{\bom}^0}
\def\0obom{{\obom}^0}

\def\0nbom{{\bom}_{n,0}}
\def\n*bom{{\bom}^*_{(n)}}

\def\LT{{\mathbb{LT}}}

\begin{document}
\title{A Mermin--Wagner theorem for Gibbs states on Lorentzian Triangulations}
\author{ M. Kelbert$^{1,3}$ \and Yu. Suhov~$^{2,3}$\and A. Yambartsev~$^{3}$}
\vspace{1mm}

\maketitle {\footnotesize
\noindent $^{1}$ Department of Mathematics, Swansea University, UK\\
E-mail: M.Kelbert@swansea.ac.uk

\noindent $^2$ Statistical Laboratory, DPMMS, University of Cambridge, UK\\
E-mail: I.M.Soukhov@statslab.cam.ac.uk

\noindent $^3$ Department of Statistics, Institute of Mathematics
and Statistics, \\ University of S\~ao Paulo, Brazil.\\
E-mail: yambar@ime.usp.br }

\begin{abstract}
We establish a Mermin--Wagner type theorem for Gibbs states on infinite random Lorentzian
triangulations (LT) arising in models of quantum gravity. Such a triangulation is
naturally related to the distribution $\sf P$ of a critical Galton--Watson tree, conditional upon non-extinction. At the vertices of the
triangles we place classical spins taking values in a torus
$M$ of dimension $d$, with a given group action of a torus ${\tt G}$ of dimension $d'\leq d$. In the main
body of the paper we assume that the spins interact via a two-body nearest-neighbor
potential $U(x,y)$ invariant under the action of ${\tt G}$. We analyze quenched Gibbs measures
generated by $U$ and prove that, for $\sf P$-almost all Lorentzian triangulations, every
such Gibbs measure is ${\tt G}$-invariant, which means the absence of spontaneous
continuous symmetry-breaking. 
\end{abstract}

\section{Introduction}

The goal of this paper is to establish a Mermin-Wagner type result (cf. \cite{MW}) for a spin system
on a random graph generated by a {\it causal dynamical Lorentzian triangulation} (CDLT for short).
The model of a CDLT was introduced in an attempt to define a gravitational path integral in
a theory of quantum gravity. See \cite{L} for a review of the relevant literature; for
a rigorous mathematical background of the model, cf. \cite{myz}.
More precisely, we analyze a spin system on a random 2D graph $T$ sampled from some natural ``uniform"
measure corresponding to a critical regime (see below). A Gibbs random field corresponding to a
given interaction potential is considered on graph $T$, giving rise to a quenched semi-direct product
measure.

We address the question of whether a continuous symmetry
assumed for the interaction potential will be inherited by an infinite-volume Gibbsian state.
It had been proved that the Hausdorff dimension of the critical CDLT equals 1 a.s. \cite{syz}. This
means that the emerging graph is essentially an infinite 2D lattice. Consequently, one can expect
the absence of spontaneous breaking of continuous symmetry. To prove this fact rigorously, we
apply techniques developed in the papers \cite{FP}, \cite{P}, \cite{KS}.

\section{Basic definitions. The main result}

\subsection{Critical Lorentzian triangulations}

We work with so-called rooted infinite CDLTs in a cylinder ${\sf C}=S\times [0,\infty)$,
where $S$ stands for a unit circle. Physically speaking, this is a $(1+1)$-type system (one
spatial and one temporal dimension). In a more realistic $(3+1)$-type case, a Mermin--Wagner
type result would have been surprising.

\begin{definition}
Consider a connected graph $G$ with countably many vertices, $V(G)$, embedded in cylinder ${\sf C}$.
A face of $G$ is a connected component of ${\sf C} \setminus G$. We say that $G$
determines a CDLT $T$ of $\sf C$ if
\begin{enumerate}
\item[(i)] all vertices of $G$ lie in the circles $S\times \{ j\}, j\in \mathbb N=\{0,1, \dots\}$;

\item[(ii)] each face of $G$ is a triangle; 

\item[(iii)] each face lies in some strip $S\times [j,j+1]$, $j\in \mathbb N$,
and has one vertex on one of the circles $S\times\{j\}, S\times\{j+1\}$ and two vertices
(together with corresponding edge) lying on the other circle;

\item[(iv)] the number of edges on circle $S\times \{j\}$ is positive for any $j\in \mathbb N$.
\end{enumerate}
\end{definition}
\noindent
It has to be mentioned that some care is needed here when one defines a triangle, due to self-loops
and multiple edges; cf. \cite{KY}.
We will say that the vertices lying in the circle $v\in S\times\{j\}$ belong to the
$j-$th layer of the CDLT. Next, define the size of a face as the number of edges incident
to it, with the convention that an edge incident to the same face on two sides
counts for two. We then call a face of size 3 (or $3$-sided face) a \emph{triangle}.

Note that
two vertices of a triangle on the same circle, say $S\times\{j\}$, may coincide with each other (in
this case the corresponding edge stretches over the whole circle $S\times\{j\}$, i.e., forms a loop).
Such a situation occurs, in particular, on the circle $S\times\{0\}$. In this paper we consider
only the case where the number of edges on the zero-level circle $S
\times \{0\}$ is equal to 1. This is a technical assumption made for simplifying the argument.

\begin{definition}
The root in a CDLT $T$ is a triangle $t$ of $T$, called
the root face, with the anti-clockwise ordering of its vertices $(x,y,z)$, where
$x$ and $y$ (and hence the edge $(x,y)$) lie in $S\times\{0\}$ (and $x$ coincides with $y$) whereas
$z$ belongs to $S\times\{1\}$.
Vertex $x$ is referred to as the root vertex. A CDLT with a root is called rooted.
\end{definition}

\begin{definition}
Two rooted CDLTs, $T$ and $T'$, are
equivalent if (i) $T$ and $T'$ are embeddings $i_T,\ i_{T'}$ of the same graph $G$, (ii) there
exists a homeomorphism $h:\sf C \to \sf C$ such that (ii1) $hi_T=i_{T'}$, (ii2)
$h$ transforms each circle $S\times \{j\}, j\in \mathbb N$ to itself and (ii3) $h$
takes the root of $T$ to the root of $T'$.
\end{definition}

In what follows, rooted CDLTs are considered up to the above equivalence.

In the same way we also can define a CDLT of a
cylinder ${\sf C}_N=S\times [0,N].$ Let $\LT_N$ and
$\LT_{\infty}$ denote the set of CDLTs with
support ${\sf C}_N$ and ${\sf C}$ correspondingly.

\subsection{ Tree parametrization of Lorentzian triangulations}

\begin{figure}[ht]
	\centering
	\includegraphics[width=4.5in]{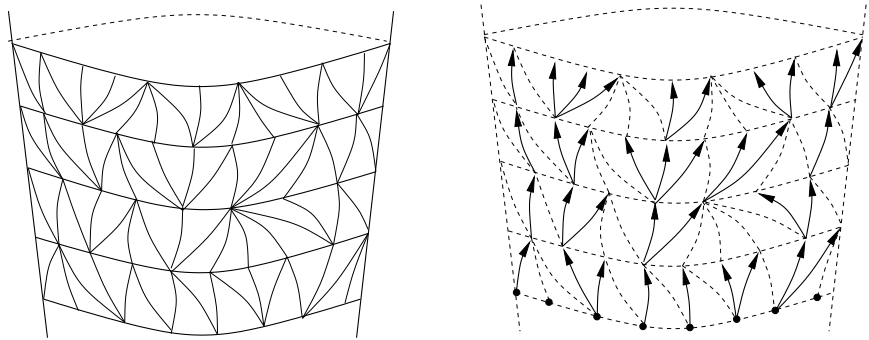}
	\caption{Tree parametrization}
	\label{fig.lt-tree}
\end{figure}

Given a CDLT $T\in\LT_N$,
define the subgraph $\tau \subset T$ by removing, for each vertex $v\in T$, the leftmost edge
going from $v$ upwards and discarding all horizontal edges, see Figure~\ref{fig.lt-tree}. The graph $\tau$ 
is formed by all the remaining edges of $T$.
The graph thus obtained is a spanning tree of $T$.
Moreover, if one associates with each vertex of $\tau$ its height in $T$
then $T$ can be completely reconstructed when we know $\tau$ (cf. \cite{KY} and references therein).
We call $\tau$ the {\it tree parametrization} of $T$.

In a similar way we can show that there exists a one-to-one bijection ${\sf m}$ between
the set $\LT_\infty$ and the set of infinite planar rooted trees $\mathcal T_\infty$:
$$
{\sf m}: \mathcal T_\infty \to \LT_\infty .$$
We will use the same symbol ${\sf m}$ for the bijection $\mathcal T_N \to \LT_N$
where $\mathcal T_N$ is the set of all planted rooted trees of height $N$.

Owing to the tree-parametrization, we are able to identify a measure ${\sf P}^{lt}_N$ on CDLTs $\LT_N$ as induced
by a measure defined on trees $\mathcal T_N$. Indeed, let ${\sf P}^{tree}_N$ be a probability measure
on $\mathcal T_N$. Then the measure ${\sf P}^{lt}_N$ on $\LT_N$ is defined by
$$
{\sf P}^{lt}_N ( {\sf m}(\tau) ) =  {\sf P}^{tree}_N ( \tau ), \mbox{ for any }\tau \in \mathcal T_N.
$$
Furthermore, if ${\sf P}^{tree}$ is a probability measure on $\mathcal T_\infty$ then the measure
${\sf P}^{lt}$ on $\LT_\infty$ is defined as
$$
{\sf P}^{lt} ( {\sf m}(A) ) =  {\sf P}^{tree}( A ), \mbox{ for any }A \subset \mathcal T_\infty.
$$
For formal details of this construction, see \cite{DJW}. In what follows we omit indices in the notation for measure $\sf P$.

To construct a measure on $\mathcal T_\infty$, we let $\nu = \{p_k\}$ be the offspring
distribution on $k\in \mathbb N = \{0,1,\dots \}$, with mean 1. Next, let $(\xi_t)_{t\ge 1}$ be the
corresponding critical Galton--Watson (GW) process. Considered under the condition that it never dies,
process $(\xi_t)$ turns into the so-called size-biased process $(k_t)$. In our setting, $k_t$ stands
for the number
of vertices on the circle $S\times \{t\}$. We refer the reader to \cite{Pem} for the detailed description
of such branching processes.

It is known that the size-biased process under consideration is supported on the subset
$\mathcal S$ of $\mathcal T_\infty$ formed by single-spine trees. A single-spine tree
consists of a single infinite linear chain $s_0, s_1, \dots $,  (here $s_0$ is the root vertex of a tree)
called the spine, to each vertex $s_j$ of which there
is attached a finite random tree with their root at $s_j$. Next, the
generating function for the branching number at each vertex $s_j$
is $f'(x)$ where $f(x)$ is the generating function of the offspring distribution $\nu$.
Moreover, the individual branches are independently and identically
distributed in accordance with the original critical GW process \cite{Pem}.
The distribution of the size-biased process $(k_t)$ generates the measure $\sf P$ as above.

Let $\sigma^2$ stand for the variance of offspring distribution $\nu$. We have
\begin{equation}\label{mart1}
{\sf E} (k_t \mid k_{t-1}) = k_{t-1} + \sigma^2.
\end{equation}
Indeed, let $\tilde \nu = \{\tilde p_k\}$ be the size biased offspring distribution:
$\tilde p_k = k p_k$. The distribution of $k_t$ conditioned on the value $k_{t-1}$ is
constructed as follows (cf. \cite{Pem}). We choose uniformly one particle from $k_{t-1}$
particles and generate the number of its descendants according to the distribution $\tilde\nu$ with
mean $\sigma^2 +1$. The descendants of other $k_{t-1}-1$ particles are generated by
the distribution $\nu$. This explain the relation (\ref{mart1}).

\subsection{Gibbs measure on Lorentzian triangulations}

Let $M$ be a $d$-dimensional torus with a flat metric $\rho$. Given a CDLT $T \in \LT_\infty$, 
the configuration space of the
spin system over $T$ is $X^{V(T)}$ where $V(T)\subset {\sf C}$ is
a collection of vertices in $T$. In the main body of the paper
we assume that spins interact via a two-body nearest-neighbor
potential $U: (x,y)\in M\times M \mapsto \mathbb R$. Nearest-neighbor
means that the spins are attached to endpoints of the same edge
of the triangulation; see below.
Let $\tt G$ be a torus of dimension $d'\leq d$ and assume $\tt G$ acts as
a group on $M$ preserving
metric $\rho$. This action is extended to $M^A$ where $A\subseteq V(T)$:
\begin{equation}\label{Gacts}
(g*\bar x_A)_v = g*x_v; \ v \in A, \ \ \bar{x}_A =\{x_v,v\in A\}\in M^A, \ g \in \tt G.
\end{equation}
Here $A\subset V(T)$ stands for a finite collection of vertices in $T$.
Next, let $\mu$ be a $\tt G$-invariant finite measure
on $M$.


We adopt the following assumptions upon $U$ (cf. \cite{DS1}).
\begin{enumerate}
\item[\bf A] {\it Invariance.} For any $g\in \tt G$ and any $x,y\in M$ $$ U(g*x, g*y) = U(x,y).$$

\item[\bf B] {\it Differentiability.} For any $x,y\in M$, there exist continuous second derivatives
$\nabla_x\nabla_y U(x,y)$.
\end{enumerate}

Given configurations $\bar x_A=\{x_v,v\in A\}\in M^{A}$ and $\bar x_{V(T)\setminus A}=\{x_v,v\in V(T)
\setminus A\}\in M^{V(T)\setminus A}$, the energy
$H(\bar x_A|\bar x_{V(T)\setminus A})$ of $\bar x_A$ in the external potential field generated by
$\bar x_{V(T)\setminus A}$ is defined by
$$
H(\bar x_A|\bar x_{V(T)\setminus A})=\sum_{\langle v, v'\rangle\in A\times A} U(x_v, x_{v'})
+\sum_{\langle v, v'\rangle\in A\times (V(T)\setminus A)} U(x_v, x_{v'}).
$$
where we use a standard notation $\langle v,v'\rangle$ for a nearest-neighbor pair. The conditional Gibbs probability density in volume $A$ with the boundary condition $x_{V(T)\setminus A}$
is defined by
$$
P( \bar x_A \mid \bar x_{V(T)\setminus A} ) = \frac{ \exp\{ - H (\bar x_A \mid \bar x_{V(T)\setminus A})
\} }{ \int \exp\{ - H ({\bar x}_A^* \mid \bar x_{V(T)\setminus A} ) \} \mu_A (d{\bar x}_A^*) },
$$
where $\mu_A$ is the product-measure on $M^A$.
A Gibbs measure ${\mathcal P}$ on $M^{V(T)}$ with potential $U$ is determined by the property
that $\forall$ finite $A\subset V(T)$, the conditional density on $M^A$ generated by ${\mathcal P}$
coincides with $P( \bar x_A \mid \bar x_{V(T)\setminus A} )$ for ${\mathcal P}$-a.s.
$\bar x_{V(T)\setminus A}\in M^{V(T)\setminus A}$ and $\mu_A$-a.s. $\bar x_A\in M^A$. Let $M^T = M^{V(T)}$ stands for the space of all configurations of spins equipped with the $\sigma$-algebra $\mathcal A_T$.

The main result of this note is
\begin{theorem}\label{main}
Assume that the offspring distribution $\nu$ has the mean 1 with finite second moment. Let $\sf P$
be the corresponding size-biased Galton-Watson tree distribution.

If the potential $U$ satisfies assumptions ${\bf A} - {\bf B}$, then every Gibbs state $\mathcal P$
with this potential is a $\tt G$-invariant measure on the space $(M^T, \mathcal A_T)$ for $\sf P$-a.a.
{\rm CLDT} $T \in \LT_\infty$.
\end{theorem}

\section{The proof: the Fr\"ohlich--Pfister argument}

The proof is based on techniques developed in \cite{FP}, \cite{P}, \cite{KS}.
First, we establish the following upper bound for the number of vertices $k_t$ on circle
$S\times\{t\}$ under the measure $\sf P$:
$\exists$ a constant $C$ (depending on a realization $T$) such that
\begin{equation}\label{ub1}
k_t \le C t \ln ^{\frac{1}{2} + \varepsilon} t, \ t=1,2,\ldots \ \ {\sf P}-a.s.
\end{equation}
In order to verify (\ref{ub1}), we note that, according to (\ref{mart1}) the process $\tilde k_t
= k_t - t\sigma^2$ is a martingale:
\begin{equation*}
{\sf E} (\tilde k_{t} \mid \tilde k_{t-1}) = {\sf E} (k_{t} \mid k_{t-1}) - t \sigma^2 =  k_{t-1} - (t-1) \sigma^2 = \tilde k_{t-1}.
\end{equation*}
It implies, in particular, that the series
$$
B_n = \sum_{t=1}^n \frac{\tilde k_{t} - \tilde k_{t-1}}{a_t}
$$
is a martingale for any sequence of numbers $\{ a_t\}, t=1,2, \dots$. Hence, we have:
$$\frac{1}{a_n}{\tilde k}_n=\frac{1}{a_n}\sum\limits_{t=1}^na_k(B_t-B_{t-1}).$$
Moreover, we can show that for some $C_1, C_2 >0$
$${\sf E}\big[(k_t-k_{t-1}-\sigma^2)^2|k_{t-1}\big]\leq C_1+ C_2k_{t-1}.$$
This follows from the representation $$k_t-k_{t-1}-\sigma^2=\xi_0-{\sf E}\xi_0+
\sum\limits_{i=1}^{k_{t-1}-1}(\xi_i-{\sf E}\xi_i),$$ where $\xi_0$ has the distribution
${\tilde\nu}$ and IID RVs $\xi_i, i=1,\ldots , k_{t-1}-1$ are distributed according
to $\nu$. Now the martingale property yields
\begin{equation}\label{ser1}
{\sf E} B_n^2 = \sum_{t=1}^n \frac{C_1 + C_2 {\sf E} k_{t-1}}{a_t^2} <
\sum_{t=1}^\infty \frac{C_1^\prime + C_2^\prime t}{a_t^2}.\end{equation}
This series converges if we choose $a_t = t (\ln t)^{\frac{1}{2} + \varepsilon}$, for any
$\varepsilon>0$. Thus, by Kroneker's lemma we have
\begin{equation}\label{as1}
\frac{k_t}{ t (\ln t)^{\frac{1}{2} + \varepsilon}} \to 0, \ \ {\sf P}\mbox{-a.s.}
\end{equation}
That proves the bound (\ref{ub1}).

Set $A\subset V(T_r)$ where $T_r$ denotes the union of the first
$r$ layers of $T$. As before $k_t$ stands for the number of vertices in $T$
lying in the layer $t$.
Let us identify $g\in \tt G$ with the vector of angles ${\underline\theta}$ and for any $n> r+1$ define the gauge action $g_n(v)\equiv g, v\in T_{r+1}$,
$g_n(v)\equiv e, v\in T\setminus T_n$. Here $e$ denotes the unit element of the group $\tt G$. On the layer $j, r+1<j<n$, we set
\begin{equation}
g_n(v)= {\underline\theta}\frac{1}{Q(n-r)}\sum\limits_{t=j+1-r}^{n-j}\frac{1}{t\ln t}, \mbox{ where }
Q(n-r)=\sum\limits_{t=2}^{n-r}\frac{1}{t\ln t}.
\label{gauge}\end{equation}

Our aim is to check the identity $\mathcal P(A) = \mathcal P(g*A)$. Using technique developed in \cite{KS} (following \cite{FP}, \cite{P}), our task
is reduced to checking the convergence of the series as $n\to\infty$
\begin{equation}\label{phi}
\phi = \sum_{ \langle v, v' \rangle} ( g_n(v) - g_n(v'))^2 \le \frac{|{\underline \theta}|^2}{ \ln\ln(n-r)} \sum_{t=2}^{n-r}
\frac{E_{t,t+1}}{ t^2 \ln^2 t},
\end{equation}
where $E_{t,t+1}$ is the number of edges that connect the vertices of $T$ from
two different levels: $S\times\{t\}$ and $S\times\{t+1\}$. By the construction of
the triangulation, $E_{t,t+1} = k_t + k_{t+1}$. Thus,
$$
\sum_{t=3}^{n-r}
\frac{E_{t,t+1}}{ t^2 \ln^2 t} = \sum_{t=3}^{n-r}
\frac{ k_t + k_{t+1} }{ t^2 \ln^2 t} < Const \sum_{t\ge 2} \frac{k_{t}}{t^2 \ln^2 t}
$$
By (\ref{as1}) the last series is
\begin{equation} \label{as2}
\sum_{t\ge 2} \frac{k_{t}}{t^2 \ln^2 t} < \infty \ \ {\sf P}-\mbox{a.s.}
\end{equation}
Hence, the sum $\phi$ in (\ref{phi}) goes to 0 when $n \to \infty$. This finishes the
proof of the theorem. $\Box$

\section{The case of the long range interaction}

The above model can be naturally extended to the case of long range interaction.
Namely, we define the energy $H(\bar x_{T_N} \mid \bar x_{T_N^c} )$
of a spin configuration $\bar x_{T_N}\in M^{T_N}$ in the external
potential field induced by a configuration $\bar x_{T_N^c}\in M^{T_N^c}$
(where $T_N^c=V(T)\setminus T_N$) as follows:
\begin{equation}\label{HAM1}
H(\bar x_{T_N} \mid \bar x_{T_N^c} ) =\sum_{(v, v')\in T_N\times T_N}
U_{v,v'}(x_v, x_{v'})
+ \sum_{(v,v')\in T_N\times T_N^c} U_{v,v'}(x_v, x_{v'}).
\end{equation}
Assume the following uniform upper bound on the interaction potentials $U_{v,v'}(x_v, x_{v'})$:
$$
|U_{v,v'}(x, x')| \le J(d(v,v')), \ \ x,x'\in M.
$$
Here $d(v,v')$ stands for the distance on graph $T$. Suppose the majorizing function $J$ satisfies, $\sf P$-a.s., the following properties:
$$
\sup_{v \in \Lambda}\,\left[ \sum_{v' \in \Lambda}  J(d(v,v')) d^2(v,v')\right]
< \infty $$
and
$$
\sup_{v \in \Lambda}\,\left[\sum_{v' \in \Lambda}  J(d(v,v')) {\bf 1}(d^2(v,v') \ge L)\right]
\to 0, \mbox{ when } L \to \infty .$$
Then, as has been shown in \cite{KS}, the assertion of Theorem 1 remains true.
Using the a.s. estimation (\ref{ub1}), the above conditions are satisfied if, for example,
$$
J(r) \le \Bigl( \frac{1}{r \ln r} \Bigr)^3.
$$

\subsection*{Acknowledgments}
This work was supported by FAPESP 2012/04372-7. M.K. thanks FAPESP 2011/20133-0 and thanks NUMEC for kind hospitality. The work of A.Y. was partially supported CNPq 308510/2010-0.


\vskip 3 truecm
\begin{thebibliography}{99}
\bibitem{myz}
V.~Malyshev, A.~Yambartsev, and A.~Zamyatin, \emph{Two-dimensional {L}orentzian
  models}, Mosc. Math. J. \textbf{1} (2001), no.~3, 439--456, 472.
  MR1877603 (2002j:82055)

\bibitem{syz}
V.~Sisko, A.~Yambartsev, and S.~Zohren, \emph{A note on weak convergence results for uniform infinite causal triangulations}, Submitted Markov Processes and Related Fields, 2011.



\bibitem{DS1}
R. L. Dobrushin and S. B. Shlosman.  Absence of breakdown of
continuous  symmetry
in two-dimensional models of statistical physics. {\it Commun. Math. Phys.},
{\bf 42}, 1975, 30-40

\bibitem{DJW} B. Durhuus, T. Jonsson, and J. F. Wheater. The spectral dimension of generic trees.
{\it J. Stat. Phys.} {\bf 128}, 2006, 1237-1260.

\bibitem{FP}
J. Fr\"ohlich and C. Pfister. On the absence of spontaneous symmetry breaking
and of crystalline ordering in two-dimensional systems. {\it Commun.
Math. Phys.}, {\bf 81}, 1981, 277--298

\bibitem{KS}
M.Kelbert and Y.Suhov. A Mermin--Wagner theorem for quantum Gibbs states on 2D graphs, I.
arXiv:1206.1229

\bibitem{KY}
M.Krikun and A.~Yambartsev. Phase transition for the Ising model on the critical
Lorentzian triangulation. Journ Statist. Phys., {\bf 148}, 2012, 422--439

\bibitem{L}
R. Loll, J. Ambjorn, and J. Jurkiewicz. {\it The universe from scratch}, Contemporary Physics, {\bf 47} (2006), 103-117.

\bibitem{Pem}
R. Lyons, R. Pemantle and Y. Peres. Conceptual proofs of $L \log L$ criteria for mean behavior of
branching processes. {\it The Annals of Probability}. 1995, Vol. 23, No 3, 1125--1138.

\bibitem{MW}
N.D. Mermin, H.  Wagner, H. Absence of ferromagnetism or antiferromagnetism
in one- or two-dimensional isotropic Heisenberg models. {\it Phys. Rev.
Lett.}, {\bf 17}  1966, 1133--1136.


\bibitem{P}
C.-E. Pfister.
On the symmetry of the Gibbs states in two-dimensional lattice systems.
{\it Commun. Math. Phys.}, {\bf 79} (1981), 181--188

\end{thebibliography}
\end{document}